# Crystal plasticity analysis of deformation anisotropy of lamellar TiAl alloy: 3D microstructure-based modelling and *in-situ* micro-compression

Liu Chen [a,b*], Thomas Edward James Edwards[c,1], Fabio Di Gioacchino[c,2], William John Clegg[c], Fionn P.E. Dunne[b], Minh-Son Pham[b]

[a] Materials Evaluation Centre for Aeronautical and Aeroengine Applications, Beijing Institute of Aeronautical Materials, Beijing 100095, China

[b] Department of Materials, Royal School of Mines, Imperial College London, London SW7 2AZ, UK

[c] Department of Materials Science and Metallurgy, 27 Charles Babbage Rd, University of Cambridge, Cambridge CB3 0FS, UK

[1] Present address: Laboratory for Mechanics of Materials and Nanostructures, Swiss Federal Laboratories for Materials Science and Technology (EMPA), Feuerwerkerstrasse 39, 3602, Thun, Switzerland

[2] Present address: ASPPRC, Dept. Metallurgical and Materials Engineering, Colorado School of Mines, 1500 Illinois St, Golden CO, US

[*] *Corresponding author. Email contact: leo.chern@yahoo.com*

**Abstract:**

Detailed microstructure characterisation and *in-situ* micropillar compression were coupled with crystal plasticity-based finite element modelling (CP-FEM) to study the micro-mechanisms of plastic anisotropy in lamellar TiAl alloys. The consideration of microstructure in both simulation and *in-situ* experiments enables in-depth understanding of micro-mechanisms responsible for the highly anisotropic deformation response of TiAl on the intra-lamella and inter-lamella scales. This study focuses on two specific configurations of $\gamma/\alpha_2$ lamellar microstructure with the $\gamma/\alpha_2$ interfaces being aligned $25^o$ and $55^o$ to the loading direction. Microstructure-based CP-FEM shows that longituginal slip of super and ordinary dislocations are most responsible for the plastic anisotropy in the $25^o$ micropillar while the anisotropy of the $55^o$ micropillar is due to longitudinal superdislocations and longitudinal twins. In addition, transversal superdislocations were more active, making the deformation in the $25^o$ micropillar less localised than that in the $55^o$ micropillar. Moreover, the CP-FEM model successfully predicted substantial build-up of internal stresses at $\gamma/\alpha_2$ interfaces, which is believed to be detrimental to the ductility in TiAl. However, as evidenced by the model, the detrimental internal stresses can be significantly relieved by the activation of transverse deformation twinning, suggesting that the ductility of TiAl can be improved by promoting transverse twins.

*Key words:* Titanium aluminide; Lamellar; Crystal plasticity; Anisotropy; Twinning

# 1. Introduction

Titanium aluminide (TiAl) alloys have a density approximately half that of nickel-based superalloys [1] and specific strength that is comparable with nickel-based alloys up to temperatures as high as 700 ºC [2], resulting in a potentially wide application of the alloys in aero-engines [3, 4]. However, the application of TiAl alloys is limited by their low ductility, which is mainly due to the anisotropy of constituent microstructure in TiAl, and insufficient understanding of the microstructure-property relationships. In-depth understanding of the relationships would lead to wider applications of these lightweight alloys in aerospace, leading to the weight reduction, thereby, increase in the fuel efficiency and decrease in the $CO_2$ emissions. Although there have been significant efforts in understanding the relationships between chemical composition, microstructure and mechanical properties of TiAl alloys, in particular regarding the TiAl lamellar microstructure [1, 5-7], there have been no detailed studies in which both *in-situ* mechanical testing and microstructure-based modelling are carried out to provide more insights into the direct relationships between constituent microstructures and mechanical properties. This present study aims to integrate modelling and *in-situ* tests to seek possible solutions to the low ductility of TiAl.

The poor ductility of lamellar TiAl alloys is partly due to its strong plastic anisotropy because of the respective anisotropy of constituent phases. Lamellar TiAl alloys consist of two phases: $\gamma$–face centred tetragonal (f.c.t.) and $\alpha_2$–hexagonal close packed phases. The $\gamma$ phase has an ordered arrangement of atoms. On each close packed (111) plane, only one slip direction, say [$\bar{1}$10], has the same atom type, while along the other two slip directions, the atom types alter periodically. The $1/2[\bar{1}10]$ dislocations are named as ordinary dislocations (characterized by a zero component along $c$ axis), while the $[10\bar{1}]$ and $[01\bar{1}]$ dislocations are termed superdislocations. Therefore, the $\gamma$ phase has twelve potential slip systems comprising four ordinary and eight super dislocation slip. It also has four twinning systems [1]. Different slip types (ordinary *vs.* super dislocation slip) and twinning generally have different critical resolved shear stress (CRSS) and hardening rates, making the $\gamma$ highly plastically anisotropic [8]. The hexagonal $\alpha_2$ phase has three families of slip: prismatic, basal and pyramidal systems whose initial CRSS ratio is about 1:3:9 in off-stoichiometric alloys [1]. In addition, the two phases are aligned to each other according to the Burgers relationship with $(111)_\gamma \parallel (0001)_{\alpha_2}$ and $\langle 1\bar{1}0 \rangle_\gamma \parallel \langle 11\bar{2}0 \rangle_{\alpha_2}$ [1]. The slip and twin systems that are transversal to the $\gamma/\alpha_2$ interfaces are not favourable, resulting in a strong dependence of plasticity on the angle between the lamellar interface and applied load [9-11].

The anisotropic elasticity and plasticity at lamella or colony scales can result in the built up of internal stresses at interfaces [12]. Such internal stresses can reach up to 80% of the yield stress in TiAl alloys [13]. The internal stresses are generally not desirable for the TiAl alloys since they can cause crack nucleation and accelerate crack propagation, resulting in the deterioration in mechanical performance. Recent research has demonstrated that internal stress leads to cracks and deteriorates both the strength and ductility [14]. Because internal stress inherently relates to microstructures of constituent phases, it is important to study the relationships between microstructure and deformation in TiAl alloys. There have been a number of studies to understand such relationships via *ex-situ* and *in-situ* mechanical testing and electron microscopy (EM) [6, 15, 16]. However, complex microstructures make it difficult to interpret experimentally observed behaviour.

Crystal plasticity finite element modelling (CP-FEM) proves to be useful to understand the complicated mechanical behaviours of titanium alloys [17-20]. 2D models were used firstly by Kad *et al.* in TiAl alloy where the $\gamma/\alpha_2$ lamellar structure was simply treated as a mixed composite [21, 22]. Schlögl and Fischer extended the 2D structure model to 3D construction of constituent phases, but without detailed incorporation of superdislocation slip [23]. Lebensohn *et al.* further extended the modelling effort by including the twelve dislocation slip and four twinning systems of $\gamma$ lamellae and three deformation modes (*i.e.* longitudinal, mixed and transversal) in their visco-plastic self-consistent (VPSC) framework. However, VPSC cannot accurately account for plastic incompatibility across lamellae as it simplifies the interaction between grains and phases [24]. Based on the work of Lebensohn *et al.*, Werwer and Cornec incorporated $\alpha_2$ lamellae and six $\gamma$ variants in a representative volume element, and used a periodic boundary condition to simulate the mechanical performance of TiAl [25, 26]. In addition, Grujicic *et al.* calibrated the crystal-plasticity parameters based on the compression experiments of single-phase pillars, and showed good agreement with the experimental observation [27]. Zambaldi *et al.* performed indentation tests to study the intra-phase behaviour and identified the slip parameters via CP-FEM [28]. Furthermore, Schnabel *et al*. utilized defect-based crystal plasticity model to extract the respective contribution of different boundaries to the macroscopic yielding [29, 30].

CP-FEM modelling that explicitly accounts for geometry and spatial distribution of lamellae is computationally expensive [31]. To address this issue, either the representative elementary volume with spatially distributed lamellae [23, 25, 29, 30, 32], or a homogenisation approach treating lamellar colony as a homogenous entity [31, 33] was used in previous works. Model parameters were often identified on the basis of experiments on bulky polysynthetically twinned (PST) crystals with off-stoichiometric chemistries [28, 33], which were not the same as the composition of TiAl alloys used in aeroengine components. Recently, micromechanical testing of TiAl alloys which are commonly used in aeroengines has proved helpful to extract the intrinsic mechanical properties of TiAl [11, 28, 33], and the small-volume specimen has made it practical to include detailed microstructure in CP-FEM. Moreover, the development of methods for mapping deformation in micromechanical testing [34] has helped linking the mechanical properties to the accumulation of plastic strain in individual lamellae [16, 35]. Accurate construction of $\gamma/\alpha_2$ lamellar microstructure in CP-FEM together with *in-situ* micro-compression and strain mapping will enable a high fidelity simulation of the deformation behaviour of TiAl alloys and offer more insights into slip and twinning activities at microscopic scale, and finally provide better understandings of the complex relationship between TiAl microstructure and deformation.

In this study, two lamellar $\gamma/\alpha_2$ micropillars of a TiAl alloy (Ti4522XD) that were carefully characterised by electron microscopy to provide accurate information of microstructure for constructing microstructure FE models of $\gamma/\alpha_2$ TiAl. Subsequently, *in-situ* mechanical testing, strain mapping and CP-FEM were systematically performed to reveal the slip and twinning activities at lamella scale, and to relate such activities to the overall deformation behaviour of lamellar $\gamma/\alpha_2$ micropillars. In particular, this study includes ordinary dislocation slip, superdislocation slip and mechanical twinning in constitutive models of 3D microstructure-based CP-FEM to enable detailed studies of plastic anisotropy, internal stress evolution and strain distributions in lamellar TiAl alloys.

## 2. Experimental and modelling methods

## 2.1. Experimental procedures

The commercial TiAl alloy with chemical composition of Ti-45Al-2Nb-2Mn (at.%)-0.8 vol.% $TiB_2$ (Ti4522XD) was used in present study. The as-received alloy was in a nearly lamellar condition after casting and hot-isostatic pressing treatment, and a typical micrograph is shown in Fig. 1.

A cuboidal sample with dimensions of $2 \times 3 \times 8\ mm^3$ were cut out and mechanically polished at two adjacent surfaces, one named as top face and the other one as side face. After polishing, the two faces were scanned by back-scattering electron (BSE) imaging technique to identify which colonies have lamella perpendicular to the side face but inclined with specific angle to the pillar axis (defined as $\Phi$ in Fig. 2a). The focused Ga+ ion beam (FIB, Helios NanoLab, FEI, USA) were then used to mill the micropillar on the edge of such colonies and two or three micropillars were milled out from each colony to check the consistency between samples. The final geometry of micropillar shows $5 \times 5\ \mu m^2$ top face and aspect ratio of about 2.5, with a taper less than $2^o$. The geometry and lamellar configuration of the micropillar are schematically illustrated in Fig. 2a and b. Due to the taper geometry, the engineering stress was calculated by using the half-height cross-sectional area.

The side surface was further milled with a low 2 kV milling step to improve the surface quality for electron backscattered diffraction scanning (EBSD, Nordlys Nano, Oxford Instruments, UK), which was carried out with a step size of $40\ nm$. Thereafter, because the fine milling always causes some rounding of the corners around the milled surface, the scanned surfaces were milled again to remove a layer of about $200\sim300\ nm$ thick to restore the sharp corners. A Pt speckle was coated over one side face of each micropillar to map the inhomogeneous straining by digital image correlation (DIC, Fig. 2e and i).

Micropillar were then compressed uniaxially at room temperature using a Synton diamond punch with $10\ um$ in diameter which was installed on an Alemnis nanoindenter of within a scanning electronic microscope (SEM). The compression was controlled by displacement with a strain rate of about $1.5 \times 10^{-3}\ s^{-1}$, and unloaded immediately after about 10% strain to freeze the microstructure for EBSD observation. After compression, an additional $200\sim300\ nm$ thick layer was removed again to enable the second EBSD scan (Scheme to determine $\gamma$ variants is described in detail in our previous publication [16]).

The pre-test EBSD images of micropillars (Fig. 2c, f) were used to construct CP-FEM models and along with the post-test scanning micrographs (Fig. 2d, g), longitudinal twinning (numbered in Fig. 2c, f) and the constraint from domain boundaries (Fig. 2g) were elucidated.

Also, by using FIB sliced and thinned lift-out samples, transversal twinning inside fine lamella was characterized using transmission electron microscopy (TEM, Fig. 2h) at 200 kV with a field emission gun in a Tecnai F20 microscope. The EBSD and DIC mapping provides an *in-situ* study of deformation mechanisms of lamellar TiAl.

## 2.2. Crystal plasticity finite element modelling

### 2.2.1. Formulation of dislocation slip

The kinematics for crystal elastic-plastic deformation is on the basis of multiplicative decomposition of the deformation gradient, $\boldsymbol{F}$, into elastic and plastic parts as $\boldsymbol{F} = \boldsymbol{F^e F^p}$ [36, 37]. The plastic component of the deformation gradient, $\boldsymbol{F^p}$, can be calculated from the plastic velocity gradient $\boldsymbol{L^p}$, as $\dot{\boldsymbol{F^p}} = \boldsymbol{L^p F^p}$. Because only the activated slip systems contribute to plastic deformation, so the plastic velocity gradient $\boldsymbol{L^p}$ is calculated from the sum of shear strain rate $\dot{\gamma}^i$ on all active systems as

$$\boldsymbol{L^p} = \sum_{i=1}^{n} \dot{\gamma}^i \boldsymbol{s}^i \otimes \boldsymbol{n}^i, \qquad (2.1)$$

where $\boldsymbol{s}^i$ and $\boldsymbol{n}^i$ denote the shear direction and plane normal of the *i*th slip system.

The shear strain rate is generally related to resolved shear stress through a classical power law [38, 39] or thermally activated expressions [36]. The latter description is based on the thermal activation process of dislocation slip to account for the rate dependence in titanium alloys [40-42]. The plastic deformation occurs when dislocations pass through barriers under thermal assistance. The shear strain rate $\dot{\gamma}$ is proportional to the average gliding velocity ($v_g$) of mobile dislocations according to the Orowan equation $\dot{\gamma} = \rho_m v_g b$, where $\rho_m$ and $b$ are the density of mobile dislocations and magnitude of Burgers vector [43]. The average velocity $v_g$ is generally given by the multiplication of the frequency $v$ to overcome obstacles and the average activation distance $d$: $v_g = vd$ [44]. The escape frequency of dislocations is described by

$$v = v_{eff} exp(-\frac{\Delta G}{k_B T}), \qquad (2.2)$$

where $v_{eff}$ is the effective frequency to escape from barriers, $\Delta G$ is the change of Gibbs free energy, $k_B$ is the Boltzmann constant, and $T$ is the absolute temperature [44-46]. In the case of dislocation interaction, the effective frequency of $\frac{v_D b}{2l}$ was usually used [44, 46], where $v_D$ is the Debye frequency and $l$ is the length of a dislocation segment between two barriers which reflects the average barrier spacing. Hence the average gliding velocity of dislocations is given as

$$v_g = \frac{v_D d b}{2l} \exp\left(-\frac{\Delta G}{k_B T}\right), \qquad (2.3)$$

where the activation energy, $\Delta G$, depends on the change of Helmholtz free energy and the work done due to external applied stress. The work done is given by the multiplication of force acting on dislocation segment $l$ and the average activation distance $d$, where the force of a net stress field $\Delta \tau = \tau - \tau_c$ exerted on dislocation segment $l$ is $\Delta \tau \cdot l b$. Hence the work done is: $\Delta \tau \cdot l b d$, and the term $l b d$ is the activation volume $V$. Consequently, the change of Gibbs free energy is obtained as

$$\Delta G = \Delta F - \Delta \tau \cdot V. \qquad (2.4)$$

The shear strain rate on the $i$th slip system is given as follows with further approximation that $l \approx d$ [36, 47]:

$$\dot{\gamma}^i = \rho_m v_D {b^i}^2 \exp\left(-\frac{\Delta F}{k_B T}\right) \sinh\left[\frac{(\tau^i - \tau_c^i)V}{k_B T}\right]. \qquad (2.5)$$

Note the physical activation volume $V$ has been expressed as $V = \gamma_0 b^2 / \sqrt{\rho_{im}}$ [48], where $\gamma_0$ is a reference shear strain conjugate to the resolved shear stress in the reference state, and the $\rho_{im}$ is the density of immobile dislocations.

In addition, the strain hardening on the $i$th slip system is assumed to follow the generalised Taylor hardening rule in which the dislocation density evolves with effective plastic strain, and the boundary hardening due to lamellar interface against transversal or mixed slip is described by a Hall-Petch-like relationship, so the critical resolved shear stress can be described as:

$$\tau_c^i = \tau_{c,0}^i + \alpha^i \mu b^i \sqrt{\eta \epsilon_{eq}} + k_{HP} \lambda_{iphase}^{-\frac{1}{2}}, \qquad (2.6)$$

where $\tau_{c,0}^i$ is the initial CRSS on the $i$th slip system, $\alpha$ and $\eta$ are constants, $\mu$ is the shear modulus, $\epsilon_{eq}$ is the von Mises equivalent plastic strain, $i.e.$ $\int \sqrt{\frac{2}{3} \dot{\varepsilon}_{ij}^p \dot{\varepsilon}_{ij}^p} dt$, $k_{HP}$ is the Hall-Petch-like coefficient which is assumed to have the same value for transversal and mixed slip for simplification [24], and $\lambda_{iphase}$ is the average lamellar spacing.

### 2.2.2. Formulation of twinning in the γ phase

The mechanical twinning can be introduced as an additional deformation mechanism in the CP-FEM framework [49, 50]. By considering the following observations: (*i*) the contribution of twinning to plasticity can be treated in the same as that of dislocation slip; (*ii*) the mechanical twinning rate relates to the shear strain rate of twinning partials; (*iii*) the contribution of dislocation slip in twinned zones should be accounted for, Kalidindi [51] proposed the following expression for the plastic velocity gradient

$$\boldsymbol{L}^p = \left(1 - \sum_{j=1}^{N_{tw}} f_{tw}^j\right) \sum_{i=1}^{N_{slip}} \dot{\gamma}^i \boldsymbol{s}^i \otimes \boldsymbol{n}^i + \sum_{j=1}^{N_{tw}} \gamma_{tw} \dot{f}_{tw}^j \boldsymbol{s}^j \otimes \boldsymbol{n}^j + \sum_{j=1}^{N_{tw}} \left(f_{tw}^j \sum_{k=1}^{N_{slip}} \dot{\gamma}^k \boldsymbol{s}^k \otimes \boldsymbol{n}^k\right), \quad (2.7)$$

where the first term denotes dislocation slip in the matrix, the second term relates to twinning, and the last represents dislocation slip in twinned zones. Although spatially discrete twin formulations would give better account of twinning activities in TiAl [52], Kalidindi's mean field treatment of twinning was used thanks to its computational efficiency, in particular when considering that this present study also aims to include dislocation slip, and morphology and spatial distribution of lamellae in TiAl. Note that the twinning is treated as pseudo slip which is unidirectional and only occurs when $\tau - \tau_{c,tw} > 0$. The twinning rate $\dot{f}_{tw}^j$ on the *j*th twinning system is believed to contribute to the shear strain rate by $\dot{\gamma}_{tw}^j = \gamma_{tw} \dot{f}_{tw}^j$, where $\gamma_{tw}$ is a shear strain induced by twinning, and theoretically equals $\frac{2(c/a)^2 - 1}{\sqrt{2}(c/a)}$ in f.c.t. materials [53]. Note that the thickness of twin lamellae in TiAl is typically less than 1 micron and the twin volume fraction is generally a few percent, which depends closely on the strain magnitude, orientation and temperature [1]. If the lattice reorientation due to twinning and subsequent slip in twinned zones are not considered [23], the equation (2.7) can be simplified as

$$\boldsymbol{L}^p = \left(1 - \sum_{j=1}^{N_{tw}} f_{tw}^j\right) \sum_{i=1}^{N_{slip}} \dot{\gamma}^i \boldsymbol{s}^i \otimes \boldsymbol{n}^i + \sum_{j=1}^{N_{tw}} \gamma_{tw} \dot{f}_{tw}^j \boldsymbol{s}^j \otimes \boldsymbol{n}^j. \quad (2.8)$$

Following this approach, the mechanical twinning is treated phenomenologically as unidirectional pseudo-slip. The shear strain rate $\dot{\gamma}_{tw}$ of twinning partials is assumed to have a similar rule as dislocation slip, Eqn. 2.5. This assumption is acceptable to account for twinning in cubic alloys. For example, the mechanical twin in f.c.c. metals acts as a result of successive gliding of Shockley partials in adjacently close-packed planes [54], which involve two processes: the nucleation of twinning partials, and their gliding in successive twinning planes. However, since twinning partials have definitely different nucleation mechanisms compared to that of dislocations, the values of parameters in Eqn. 2.5 (such as physical activation volume $V$) should be different for twinning. Moreover, as twinning nuclei require at least three stacking faults [54], the unit contribution to plastic deformation needs a successive gliding of twinning partials. Therefore, an effective Burgers vector $\boldsymbol{b}_{eff}$ was introduced to describe both the influence of the physical activation volume due to the different nucleation mechanisms, and the successive gliding character of twinning partials, as $\boldsymbol{b}_{eff} = b_0 \boldsymbol{b}_{tw}$, where $\boldsymbol{b}_{tw}$ is the Burgers vector of twinning partials, and $b_0$ is a constant which needs to be identified. Hence the shear strain rate on the $j$th twinning system is

$$\dot{\gamma}_{tw}^j = \rho_m v_D b_{eff}^2 \exp\left(-\frac{\Delta F}{k_B T}\right) \sinh\left[\frac{(\tau^j - \tau_{c,tw}^j)V}{k_B T}\right]. \tag{2.9}$$

The occurrence of current mechanical twinning will reduce the volume (of matrix) that is available for subsequent twinning. This might result in an increase in the critical stress to activate subsequent twinning. It is assumed that the threshold stress $\tau_c^j$ for twinning evolves and follows a similar hardening rule as that for slip, *i.e.*,

$$\tau_{c,tw}^j = \tau_{c,0} + \alpha_{tw}^j \mu b_{eff} \sqrt{\eta_{tw} \epsilon_{eq}} + k_{HP} \lambda_\gamma^{-\frac{1}{2}}, \tag{2.10}$$

where $\tau_{c,0}$ is the initial CRSS to activate twinning, $\alpha_{tw}^j$ is a constant depending on whether twinning is longitudinal or transversal to the interface between lamellae, $\mu$ is the shear modulus, $\eta_{tw}$ is a constant which was set equal to $\eta$ in dislocation slip, $k_{HP}$ is a Hall-Petch-like coefficient and $\lambda_\gamma$ is the average spacing of $\gamma$ lamellae.

### 2.2.3. Parameter identification of $\gamma$ and $\alpha_2$ single phases

FEM models of $2 \times 2 \times 4 \; mm^3$ were used for both $\gamma$ and $\alpha_2$ single phases in accordance with experiments reported in [55, 56], which were meshed with 4563 elements. The crystal model (Section 2.2.2) was then coded into ABAQUS/user-defined finite element (UEL) with three-dimensional 20-noded quadratic (C3D20R) elements. The compressive loading was vertically along the height direction, and the bottom surfaces ($2 \times 2 \; mm^2$) were fixed during deformation.

The $\gamma$ phase has lattice constants: $a = 0.401 \; nm$ and $c/a = 1.02$, while the lattice constants of the $\alpha_2$ phase are: $a = 0.577 \; nm$ and $c/a = 0.8$ [1, 27]. Fig. 3 shows the f.c.t. crystal structure and its ordinary and super dislocation slip systems. Fig. 4 shows the hexagonal crystal structure of $\alpha_2$ phase. Three types of slip named as basal, prismatic and pyramidal are shown in Fig. 4b−d. The anisotropic elasticity for both phases is incorporated via the elastic modulus matrices: the elastic matrix of the $\gamma$ phase is $C_{11} = 190, C_{12} = 105, C_{13} = 90, C_{33} = 185, C_{44} = 120, C_{66} = 50$ [57], while that of the $\alpha_2$ phase is $C_{11} = 221, C_{12} = 71, C_{13} = 85, C_{33} = 238, C_{44} = 69, C_{66} = (C_{11} - C_{12})/2$ (all are in GPa) [58]. Shear moduli of $\gamma$ and $\alpha_2$ phases at room temperature are about 70 GPa and 57 GPa, respectively [1].

The $\gamma$ phase has twelve slip systems (including four ordinary dislocation and eight superdislocation slip systems), and four twinning systems (Table A.1). Regarding the hexagonal $\alpha_2$ phase, twelve slip systems (including three $(10\bar{1}0)[\bar{1}2\bar{1}0]$ prismatic, three $(0001)[\bar{1}2\bar{1}0]$ basal and six $(11\bar{2}1)[\bar{1}\bar{1}26]$ pyramidal slip systems) were taken into consideration and summarized in Table A.2. The Debye frequency and Boltzmann constant are known and listed in Table 1. The reference shear strain is taken to be $\gamma_0 = 6.0 \times 10^{-4}$ which has been used in literature to simulate the plastic deformation of titanium alloys [48]. The average dislocation density in TiAl alloys is typically $\sim 10^{12}\ m^{-2}$ [1]. Since the $\gamma$ phase usually accommodates more plasticity than $\alpha_2$, there should be more dislocations in $\gamma$ than in $\alpha_2$. We assume that the mobile and immobile dislocation densities in the $\gamma$ phase are $5 \times 10^{12}\ m^{-2}$ and $1 \times 10^{12}\ m^{-2}$, respectively, while in $\alpha_2$ phase are $4 \times 10^{12}\ m^{-2}$ and $1 \times 10^{12}\ m^{-2}$ (Table 1). Because TiAl has a relatively low strain rate sensitivity at room temperature [1, 55], $7.8 \times 10^{-20}$ J is used for the activation energy.

The critical resolved shear stresses for slip systems in the hardening rules of equation (2.6) or (2.10) were identified as follows. For single phase $\gamma$, the slip of ordinary dislocations is easiest, with an initial CRSS of $\tau_0$, while the superdislocation slip is harder to activate. Therefore, the initial CRSS for superdislocation slip is $k_{so} \times \tau_0$ where $k_{so}$ ($k_{so} > 1$) is a constant. Using the CP-FEM to fit the experimental stress-strain curves during compression along the [001] and [010] directions [55] (Fig. 5a), the initial CRSSs were calibrated to be 45 MPa and 55 MPa for ordinary and super dislocation slip, respectively. The initial CRSS of twinning was set to be the same as ordinary dislocations. The factor $k_{so}$ was calibrated to be 1.2. In addition, the $\alpha$ values in the hardening rule for ordinary dislocation, superdislocation and twinning partial were calibrated to be 0.55, 0.45 and 0.28, respectively. Values of 460 and 2.5 were used for $\eta$ and $b_0$, respectively.

For the $\alpha_2$ phase, the easiest slip is prismatic, and its initial CRSS is written as $\tau_p$. The initial CRSSs for basal and pyramidal slip are $k_{bp} \times \tau_p$ and $k_{pyp} \times \tau_p$, respectively. These parameters were calibrated from fitting experimental stress-strain curves [56] as shown in Fig. 6a. The $\tau_p$, $k_{bp}$ and $k_{pyp}$ values were calibrated to be 85 MPa, 2.0 and 9.5, respectively. For the hardening rule, the $\alpha$ values for prismatic, basal and pyramidal dislocations were 0.8, 1.28 and 3.2, while $\eta$ was set to 100.

Fig. 5 shows excellent agreement between experimental [55] and simulated stress – strain curves of single-phase $\gamma$ compressed along different crystallographic orientations. There is a slight variation in the yield stresses along different crystallographic orientations ranging from 200 to 240 MPa, as shown in Fig. 5a and b. The $\gamma$ crystal exhibits a CRSS ratio of 1.2 between super and ordinary dislocations, which is less than the literature value of 1.5~1.8 [59, 60] probably due to its off-stoichometric composition. In contrast with the small anisotropy in yielding, there was significant anisotropy in the hardening of $\gamma$. Fig. 5 shows that after 3% compression, the flow stress increased by about 300 MPa along the [001] orientation, but only by 80 MPa along [$\bar{2}$96] or [$\bar{2}$45]. The value of the hardening term $\alpha\mu b\sqrt{\eta}$ in the hardening equation 2.6 was calibrated based on the experimentally observed hardening behaviour in Fig. 5. 401 MPa, 661 MPa and 510 MPa were identified for the hardening term of ordinary, super dislocation slip and twinning, respectively.

Fig. 6 shows the experimental [56] and simulated stress – strain curves in $\alpha_2$ single phase crystals. The yield stresses were substantially different along different orientations. The [0001] basal orientation has the highest initial CRSS as the potential slip under this orientation is pyramidal, while the [11$\bar{2}$0] orientation is the softest. Using the FEM model with the proposed constitutive models for $\alpha_2$ single phase to fit against the curves gives (1) the values of initial CRSSs for prismatic, basal and pyramidal slip to be: 85 MPa, 170 MPa, and 808 MPa; and (2) the values of hardening $\alpha\mu b\sqrt{\eta}$ term for prismatic, basal and pyramidal slip to be: 319 MPa, 510 MPa, and 1200 MPa, respectively.

### 2.2.4. Microstructure-based CP-FEM of $\gamma/\alpha_2$ lamellar TiAl

The CP-FEM models of $\gamma/\alpha_2$ lamellar microstructure were carefully reconstructed based on experimental observation via EBSD scans of frontal surfaces [16], which were then meshed and coded into ABAQUS/UEL with C3D20R element type. The microstructure observed by EBSD scans was assumed to be identical through thickness. Fig. 7a is the front view of an undeformed micropillar, and the corresponding EBSD map is shown in Fig. 7b. Multiple $\gamma$ variants and $\alpha_2$ lamellae were coloured based on the Euler angle values (the $\phi_1 = 0$, $\Phi = 0$ and $\phi_2 = 0$ were represented by red, green and blue colour; readers are referred to ref. [16] for the definition of $\gamma$ variants). The average spacing of $\gamma$ and $\alpha_2$ lamellae measured from EBSD results was about $1\ \mu m$. The angle $\phi$ defined as the acute angle between the lamellar interface and loading direction, was used to differentiate micropillars throughout this manuscript. Fig. 7c shows the CP-FEM model with $\phi = 25^o$ (meshed with 57267 elements), in which the geometry and crystal orientation were based on Fig. 7a and b. Similarly, Fig. 7d, e and f show the SEM view, EBSD map and corresponding CP-FEM model of micropillar with $\phi = 55^o$ (meshed with 53325 elements). Because the micropillars used in experiments were very small compared to their bases, it is reasonable to assume the bases as a rigid body. Correspondingly, the nodes on the bottom face in simulation were fully fixed during simulation for both micropillar models.

Since the chemical composition and microstructure of the two $\gamma/\alpha_2$ lamellar micropillars were different from those of the single-phase pillars in Section 2.2.3, some parameters needed to be recalibrated based on fitting experimental stress-strain curves of the two $\gamma/\alpha_2$ lamellar micropillars by CP-FEM. The fitting gave $k_{so} = 1.8$ and $\tau_0 = 70$ MPa for the $\gamma$ phase. For hardening parameters, the same values identified for single phase pillars (Section 2.2.3) were used to describe the strain hardening of each individual phase in the $\gamma/\alpha_2$ lamellae. Boundary hardening associated with $\gamma/\gamma$ and $\gamma/\alpha_2$ interfaces is accounted for by $k_{HP}\lambda_{iphase}^{-\frac{1}{2}}$ in Eqn. 2.6. The value of $k_{HP}$ was 140 MPa$\sqrt{um}$ for transversal and mixed deformation mode (according to reference [61]) while $k_{HP} = 0$ MPa$\sqrt{um}$ for longitudinal deformation mode.

The definition of deformation modes (namely, longitudinal, transversal and mixed) in this study was adopted from reference [24]. In detail, the longitudinal mode has both the slip plane and direction parallel to lamellar interface, while in transversal mode both the slip plane and direction are inclined to the lamellar interface. If the slip (or twinning) plane is inclined to, but with slip (or twinning) direction parallel with, the lamellar interface, the system is of the mixed mode. This definition of deformation mode is applied to both the $\gamma$ and $\alpha_2$ phases. According to this definition and the Burgers orientation relationship between two phases, *i.e.*, $(1\bar{1}1)_\gamma//(0001)_{\alpha_2}$, the deformation modes in $\gamma$ and $\alpha_2$ lamellae were defined and given in tables A.1 and A.2. For example, the basal, prismatic and pyramidal slip in $\alpha_2$ were identified as longitudinal, mixed and transversal mode, respectively.

## 3. Results

### 3.1. Overall plastic deformation: In-situ observation *versus* CP-FEM

CP-FEM models with accurate microstructure input based on EBSD characterisation (Fig. 7) were able to describe the experimental stress − strain curves of two lamellar micropillars observed by *in-situ* mechanical tests (Fig. 8). Note the global strain was measured on the basis of the displacement of the indenter and the test date was corrected for test rig. CP-FEM also describes well the orientation dependence of mechanical properties that the $55^o$ micropillar has a relatively lower yield stress but higher hardening rate than the $25^o$ micropillar. Most importantly, the sites of localised deformation bands are well predicted by CP-FEM. Fig. 9 shows excellent agreement in the location and degree of plastic strain accumulation in individual lamellae between experimental observations (1st and 3rd rows) and simulations (2nd and 4th rows) at different nominal strains for two micropillars. In particular, the localized deformation bands in the $25^o$ micropillar were parallel to lamellar interfaces (white arrows in the first row, Fig. 9e) which was also predicted in the CP-FEM (black arrows in the second row of Fig. 9e). For the $55^o$ micropillar, experimental observations show that the localized shear bands started developing clearly at the nominal strain of 1.5% and subsequently left distinct steps on free surfaces (as marked by arrows in Fig. 9j). Such activities of shear bands were well captured by the CP-FEM model (see strain fields obtained from CP-FEM shown in the 2nd row of Fig. 9f-g). The excellent degree of agreement between experiment and CP-FEM indicates that the microstructure-based CP-FEM is of high fidelity.

### 3.2. Slip and twinning activities

By coupling the accurate microstructure and dislocation-based constitutive laws, CP-FEM enables identification of the activities of individual slip and twinning systems and their contribution to the overall response. The activity of separate slip/twinning modes in each lamella for $25^o$ micropillar is revealed in Fig. 10. Deformation in the micropillar was mainly accommodated by longitudinal modes. For example, substantial longitudinal slip associated with ordinary dislocation $(1\bar{1}1)[110]$ in $II_M$, superdislocation $(1\bar{1}1)[011]$ in $II_M$ and $III_M$ and superdislocation $(1\bar{1}1)[10\bar{1}]$ in $I_M$ were predicted by CP-FEM (Fig. 10a, b and c). Similarly, the CP-FEM model predicted strong longitudinal twinning in $III_T$ (Fig. 10d). Although the CP-FEM indicated that transversal twinning $(\bar{1}11)[\bar{1}1\bar{2}]$ was active in $III_T$ (Fig. 10e), this mechanism was overwhelmed by longitudinal twinning (Fig. 10d). In the $\alpha_2$ lamellae, Fig. 10f shows the plastic strain due to prismatic slip on $(\bar{1}100)[\bar{1}\bar{1}20]$ slip systems that are of mixed deformation mode. However, the slip activity of $(\bar{1}100)[\bar{1}\bar{1}20]$ slip systems was rather limited.

Fig. 11 shows the simulated slip and twinning for the $55^o$ micropillar. As shown in Fig. 11a, the plastic deformation in lamella $I_M$ was dominated by longitudinal mechanical twinning, leading to the twinning-induced localisation along interfaces in lamellar TiAl. In agreement with the simulation, the EBSD scan also detected the predominant longitudinal twinning [16]. Interestingly, the simulation results predicted weak transversal twinning in the same lamella $I_M$ (Fig. 11b) which was not detected by EBSD. This difference is discussed later in Section 4.3. In addition to twinning in the $I_M$, localised deformation was also caused by longitudinal superdislocation slip (Fig. 11c). There was a gradient of accumulated strain associated with superdislocation activity: higher accumulated slip on the top right corner of the pillar (Fig. 11c). The $\alpha_2$ lamellae deformed with mixed prismatic slip which was more active in the region below the $I_M$ variant of $\gamma$ (Fig. 11d).

The contributions of longitudinal, transverse and mixed deformation modes to the total plastic strain in the two micropillars are quantified and presented in Fig. 12 (a, b are for the $25^o$ micropillar; and c, d for the $55^o$ micropillar). The CP-FEM model was able to reveal the activity of individual slip and twin systems and their contributions to the three different deformation modes (Fig. 12b and d). To compare the deformation heterogeneity because of the plastic anisotropy of individual constituent lamellae, the maximum shear stress and strain along a specific path (named the 0-1 path highlighted by a dashed line in Fig. 13a for the $25^o$ micropillar and in Fig. 13d for the $55^o$ micropillar) were extracted. Fig. 13b and c show substantial differences in maximum shear strain in lamellae A, B, C, D and E with their adjoining neighbours in the $25^o$ micropillar. Similarly, large difference in maximum shear strain of lamellae A, B and C with their neighbours in the $55^o$ micropillar. Along the paths (Fig. 13a, d), high local shear stress was usually observed in $\alpha_2$ lamellae in which corresponding shear strain was low (Fig. 13c and f versus Fig. 13b and e). The obtained insights into slip and twinning activities and their contributions to inhomogeneous deformation presented in Figs. 12 and 13 will be used to understand the mechanisms of localisation in the two micropillars and will be discussed in Sections 4.2 and 4.3.

## 4. Discussion

### 4.1. Plastic anisotropy: intra-lamellar and inter-lamellar contributions

The lamellar microstructure leads to anisotropic deformation which is often not desirable for alloys under complex stress conditions. The anisotropy of lamellar microstructure results from two sources: elasticity and plasticity. While elastic anisotropy does not significantly change with deformation, the plastic anisotropy substantially evolves with plastic deformation. By including detailed consideration of microstructure and dislocation and twinning activity,

**14** / 42

this investigation was able to reveal the contributions of the constituent microstructure at different length-scales to the overall plastic anisotropy.

The first source of plastic anisotropy is associated with the microstructure behaviour within each lamella. The intra-lamellar anisotropy results from different initial CRSSs and strain hardening rates of constituent phases. For the present TiAl alloy (Ti4522XD), the initial CRSS ratio between superdislocation and ordinary dislocation slip was estimated to be 1.8 for the $\gamma$ phase in the lamellar microstructure (see the $k_{so}$ value in Table A.1). In the hexagonal $\alpha_2$ phase, the initial CRSS ratio between prismatic, basal and pyramidal slip was identified to be $1:2:9.5$. The anisotropy of the $\gamma$ phase is less than the $\alpha_2$ because the f.c.t. structure of $\gamma$ phase is more symmetric than the hexagonal structure of $\alpha_2$ [1]. The plastic anisotropy of each phase evolves because of the hardening behaviour during deformation. The ratio of the hardening coefficient $\alpha\mu b\sqrt{\eta}$ (in Eqn. 2.6) between ordinary dislocation slip, superdislocation slip and twinning in the $\gamma$ phase was $1:1.6:1.3$. The strain hardening generally depends on complex interaction between dislocations on different slip systems [62]. The higher hardening rate is possibly due to strong junctions formed after dislocation interaction. As proposed by Greenberg *et al.* [63, 64], superdislocations tend to dissociate into non-coplanar barriers during plastic deformation so that their further gliding is restricted, resulting in higher hardening rate for superdislocation slip. The different hardening rates cause the $\gamma$ phase to become increasingly anisotropic during plastic deformation. It is similar for the $\alpha_2$ phase whose hardening ratio between prismatic, basal and pyramidal slip was $1:1.6:3.8$.

In addition to the intra-lamellar anisotropy of constituent phases, plastic anisotropy of lamellar TiAl also results from the difference in plastic deformation between lamellae, i.e., inter-lamella anisotropy. The inter-lamella anisotropy relates to the orientation of interfaces between lamellae and to the Burgers orientation relationship between them. Lamellar interfaces act as barriers to dislocation movement, making the CRSS for transversal deformation higher than that of longitudinal slip. Moreover, the Burgers orientation relationship requires <a>-type prismatic slip in $\alpha_2$ to be parallel to lamellar interfaces while hard pyramidal slip is inclined to the interfaces, restricting the slip transfer across $\gamma/\alpha_2$ interfaces, leading to the inter-lamella anisotropy that is strongly dependent on the orientation of the interfaces. It should be noted that substantial hardening associated with lamellar boundaries can exert an overall strong resistance to the mixed and transversal deformation. The strengthening associated with lamellar boundary is about 140 MPa for the currently considered micropillars whose average lamellar spacing of about 1μm. The substantial hardening due to lamellar boundaries possibly makes the non-longitudinal deformation difficult to be activated (Fig. 12), even for lamella that have high Schmid factor (0.48) for the transversal twinning such as the $\gamma$ $III_T$ lamella in Fig. 10 which showed little transversal twinning.

**4.2. Deformation localisation in lamellar microstructure**

$\gamma$ and $\alpha_2$ phases have different propensities for plastic deformation because of different CRSS values. The $\alpha_2$ lamellae are usually considered as a hard phase (i.e., more difficult to be plastically deformed) as they possess higher initial CRSSs than those of $\gamma$ (refer to Sections 2.2.3 and 2.2.4, and Fig. 9). This explains why the maximum shear strain is low; and thereby, the maximum shear stress is high in $\alpha_2$ lamellae (Fig. 13). Even in each lamella, the degree of plastic deformation along a specific crystallographic orientation varies depending on the angle between the crystal orientations with respect to the applied stress. The Schmid factor (given in Table A.3) best represents the dependence between slip/twinning systems and loading direction. The longitudinal slip of ordinary dislocation $(1\bar{1}1)[110]$ in $II_M$ was active (Fig. 10a) because their Schmid factor was high (namely, 0.36). Similarly, the longitudinal slip of superdislocation $(1\bar{1}1)[011]$ in $III_M$ (Fig. 10b) was highly active because it had a Schmid factor of 0.36. Note that the ordinary dislocation motion contributed less to plastic strain (Fig. 10a, b) because it has a small number of slip systems compared with superdislocation. Plastic deformation tends to localize on slip systems that have low CRSSs and high Schmid factors, leading to different degrees of hardening which can increase the plastic anisotropy in $\gamma$ lamellae. In addition, as discussed in Section 4.1, the angle between interfaces and loading direction strongly affects the localization: *e.g.*, $55^o$ micropillar shows more intense localization compared to that of $25^o$ micropillar (Fig. 9). Consequently, differences in the Schmid factor, CRSSs and hardening behaviour of $\gamma$ and $\alpha_2$ phases lead to highly heterogeneous deformation of lamellar TiAl which is successfully revealed by CP-FEM (Figs. 9−11). Plastic heterogeneity can lead to a negative impact on the mechanical performance of alloys, *e.g.*, internal stresses will increase substantially at interfaces between soft and hard lamellae because of plastic incompatibility between them (Fig. 13), making interfaces susceptible to crack initiation [65].

Together with *in-situ* mechanical testing, this CP-FEM study provides significant insights into the strain accumulated by individual slip and twinning activities and their contributions to the overall localisation of deformation (Fig. 12). The separate activity and independent contributions of slip and twinning shed light on underlying mechanisms that are responsible for the localisation of plasticity in TiAl. For example, longitudinal slip of superdislocations in $I_M$ is responsible for the most dominant deformation band experimentally observed in the $25^o$ micropillar (Fig. 10c). Interestingly, superdislocation slip transverse to $\gamma/\alpha_2$ interfaces in the $25^o$ micropillar was active (Fig. 12b). Such activity of transversal superdislocation rendered the localization of deformation in the pillar significantly more diffuse (Fig. 9a-e). Whereas, in the $55^o$ micropillar, both longitudinal superdislocation slip and longitudinal twinning were very active during plastic deformation (Figs 11a, c and 12c, d), leading to more severe localisation along lamellar interfaces in this pillar (Fig. 9f-j). These observations are consistent with previous results that showed longitudinal deformation is dominant in $45^o$ aligned pillar under compression, and the mixed-mode deformation activities increase for lamellae being parallel to the loading axis [66].

The degree of deformation in $\alpha_2$ lamellae is much less than that in $\gamma$ because of the high CRSSs of $\alpha_2$ especially for basal and pyramidal slip (Fig. 13). Limited plastic deformation in $\alpha_2$ makes the slip transfer across $\gamma/\alpha_2$ interfaces difficult (Fig. 10f and 11d), further contributing to deformation heterogeneity in lamellar TiAl. As a result, internal stress can be built up rapidly near the interfaces. High internal stresses near $\gamma/\alpha_2$ interfaces were predicted by CP-FEM (Fig. 13c, f). Not only were there different degrees of plastic deformation between $\gamma$ and $\alpha_2$, but also there was a substantial difference in strain/stress responses between $\gamma$ variants. This heterogeneous distribution feature for stress is similar to previous observations in titanium alloys containing lamellae, in which the load shedding has pronounced influence on their mechanical properties [32].

Sufficient plastic deformation in some $\gamma$ variants caused significant stress relief in the variants. For example, this phenomenon can be seen in the variant $III_T$ between two thin lamellae B and C: high plastic deformation in this variant (Fig. 13b) leads to low shear stress (Fig. 13c). However, just across C, plastic strain in another $\gamma$ variant (between C and D) whose plasticity was low experienced much higher shear stresses, in particular near $\gamma/\alpha_2$ interfaces.

### 4.3. Deformation twinning and its influence on mechanical properties

The simulation of mechanical twinning is in good agreement with experimental observations regarding the activity of twinning. For $25^o$ micropillar, as shown by the TEM observation (Fig. 2h), the transverse twins were aligned about $69^o$ to the lamellar interface when projected to the front face of the pillar. This angle in the 3D space can be calculated theoretically through the Euler matrix $Q_{ij}$ of $III_T$ and the twin plane normal $n'_j$ in sample coordinates for $(\bar{1}11)[\bar{1}1\bar{2}]$ transversal twinning : $n'_j = Q_{ij}n_j$. As the interface normal before compression is $\mathbf{n}_l = [cos\phi, sin\phi, 0]^T$, the projections of twinning plane and lamellar interface to the front plane are then calculated as $\mathbf{V}_{proj}^{twinPlane} = [1.2174, 1.2126, 0]$ and $\mathbf{V}_{proj}^{lamInterface} = [-0.4226, 0.9063, 0]$. Then the angle between transversal twin and lamellar interface is calculated as $70.1°$, which reasonably agrees with the experimental result ($69°$). The small deviation probably results from the distortion due to plastic deformation, i.e. the rotation of the central region of the pillar during compression, identified previously [16].

There are multiple interesting points that were revealed by the CP-FEM simulation, but were not captured by experiments. Firstly, the longitudinal twinning in $III_T$ predicted by the CP-FEM simulation (Fig. 10d) was not observed by EBSD observation. A reason for this difference is that the $III_T$ was constructed to be uniform through the thickness of the pillar model, while it is not the case in the experimental pillar, as there is an inclined domain boundary with the $II_T$ variant in the same lamellae (Fig. 2f). This causes the $III_T$ domain to be more confined in the physical pillar. This restriction probably inhibited the activation of the longitudinal twinning of $III_T$ [16, 35], explaining why it was not seen in the experiment, as concluded previously [16, 35, 67]. Secondly, although EBSD shows that there was an occurrence of longitudinal twinning in lamella $I_M$ (Fig. 2f), the CP-FEM model suggests that it was superdislocation slip (Fig. 10c) that dominated deformation due to the low Schmid factor for twinning (Table A.3, 0.25 for twinning but 0.35 for superdislocation slip). This second difference likely results from the same reason of the difference in the input microstructure. While the constructed microstructure in the CP-FEM model was through-thickness, the milling process after compression of the corresponding experimental pillar did reveal a different $\gamma$ variant (and different associated twinning) lying beneath $I_M$. Thirdly, although longitudinal twinning in lamella $I_M$ of the $55°$ micropillar was seen by both CP-FEM simulation (Fig. 11a) and EBSD scans (Fig. 2c), transverse twinning predicted by simulation (Fig. 11b) had not been detected by EBSD imaging. This difference can probably result from the assumptions that the volume of twinning is small and the contribution of slip in twinned regions is negligible (refer to equations 2.7 and 2.8). Such assumptions might not hold true for such lamellae if they underwent longitudinal twinning where the twinning volume could be significant.

The activation of twinning can offer an additional way to accommodate the plastic incompatibility across the lamellar interface, resulting in a reduction in internal stress. The increase in internal stresses due to strain localisation can lead to the decohesion of interfaces, causing the initiation of cracks. Once the crack is nucleated, it will propagate quickly along interfaces as these are easy cleavage planes [1]. This explains why cracking often occurs along lamellar boundaries in lamellar TiAl [68, 69]. The activation of mechanical twinning in $\gamma$ assists the accommodation of plastic deformation via twinning. CP-FEM shows that both longitudinal (Fig. 10d) and transversal twinning (Fig. 10e) had been activated in $III_T$ lamellae (indicated by arrows) in $25^o$ micropillar. The internal shear stresses in $III_T$ lamellae (between B and C regions in Fig. 13c) were ~230 MPa which is lower than that of ~340 MPa in $\gamma$ lamellae without twinning, *e.g.*, $\gamma$ variants near $\alpha_2$ lamellae D and E in Fig. 13c. This indicates the importance of mechanical twinning as an additional mechanism in relieving internal stresses and in accommodating the plastic incompatibility across lamellar interfaces. This finding is important for seeking a way to improve the ductility of TiAl. Because twinning would not deteriorate the high-temperature strength of TiAl [70], tailoring the chemical composition to promote deformation twinning might help to improve the ductility of TiAl while still maintaining excellent strength at elevated temperatures.

## 5. Conclusions

This study develops detailed crystal plasticity finite element models (CP-FEM) of $\gamma/\alpha_2$ lamellar TiAl micropillars based on careful EBSD characterisation. Thermally activated constitutive laws were successfully implemented in the microstructure models to account for the dislocation slip and mechanical twinning. Subsequently, CP-FEM simulations were coupled with *in-situ* deformation mapping to reveal significant insights into the complex anisotropy of lamellar TiAl alloys. The coupling helps to obtain in-depth and quantitative understandings of two multi-scale micro-mechanisms for plastic anisotropy in TiAl alloys: (*i*) the in-phase anisotropy of individual constituent phases because of the distinct differences in CRSSs and hardening rates of slip and twinning systems within individual phases; (*ii*) the anisotropy because of the different plastic deformation behaviour between phases. CP-FEM was able to reveal in significant detail the dependence of TiAl anisotropy on the orientation of lamellar $\gamma/\alpha_2$ interfaces and boundaries between $\gamma$ variants, and on the crystallographic relationships between these phases.

This study particularly focused on two specific configurations which have the $\gamma/\alpha_2$ interfaces aligned $25^o$ and $55^o$ to the loading direction. From the detailed constructure of 3D microstructure, the CP-FEM model shows that the longitudinal slip of superdislocations and ordinary dislocations are most responsible for the anisotropy in the $25^o$ micropillar while the longitudinal superdislocations and longitudinal twinnings most contribute to the anisotropy of the $55^o$ micropillar. The strong anisotropy causes the localisation of plastic deformation, in particular along lamellar interfaces, leading to the build-up of internal stresses and making such interfaces susceptible to cracking. Most interestingly, 3D microstructure-based CP-FEM successfully reveals that mechanical twinning can play an influential role on relieving detrimental internal stresses, which implies a potential toughening strategy for TiAl alloys by introduction mechanical twinning through alloying engineering.

**Acknowledgements**

Much appreciated is the strong support received from Beijing Institute of Aeronautical Materials (BIAM). The research was performed at the BIAM-Imperial Centre for Materials Characterization, Processing and Modelling at Imperial College London. L.C. would like to acknowledge the financial support from the National Natural Science Foundation of China with grant no. of 51301187. T.E.J., F. D. and W.J.C. would like to acknowledge the support of the EPSRC / Rolls-Royce Strategic Partnership (EP/M005607/1).

**Appendix A**

Table A.1 contains the 12 slip systems (including 8 superdislocation and 4 ordinary dislocation slip) and 4 twinning systems of $\gamma$ phase as well as corresponding deformation modes, while table A.2 tabulates the slip systems (including 3 basal, 3 prismatic and 6 pyramidal slip) and deformation modes of $\alpha_2$ phase. Table A.3 lists the maximum Schmid factors of $\gamma$ variants for different deformation mechanisms in lamellar micropillars.

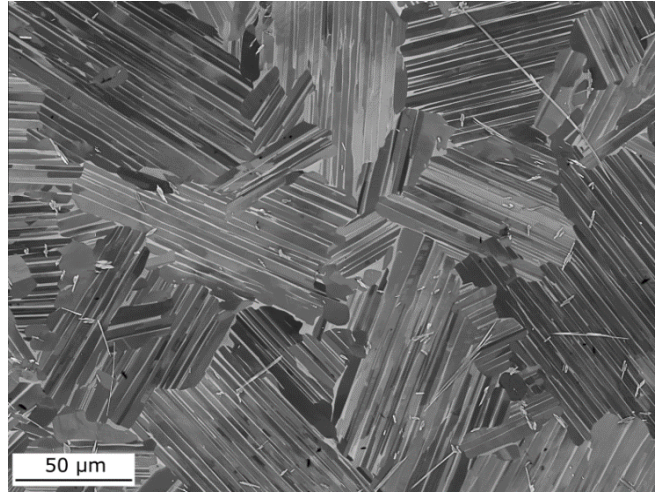

Fig. 1. Backscattered electron image showing the microstructure of the as-received material following centrifugal casting and hot-isostatic pressing.

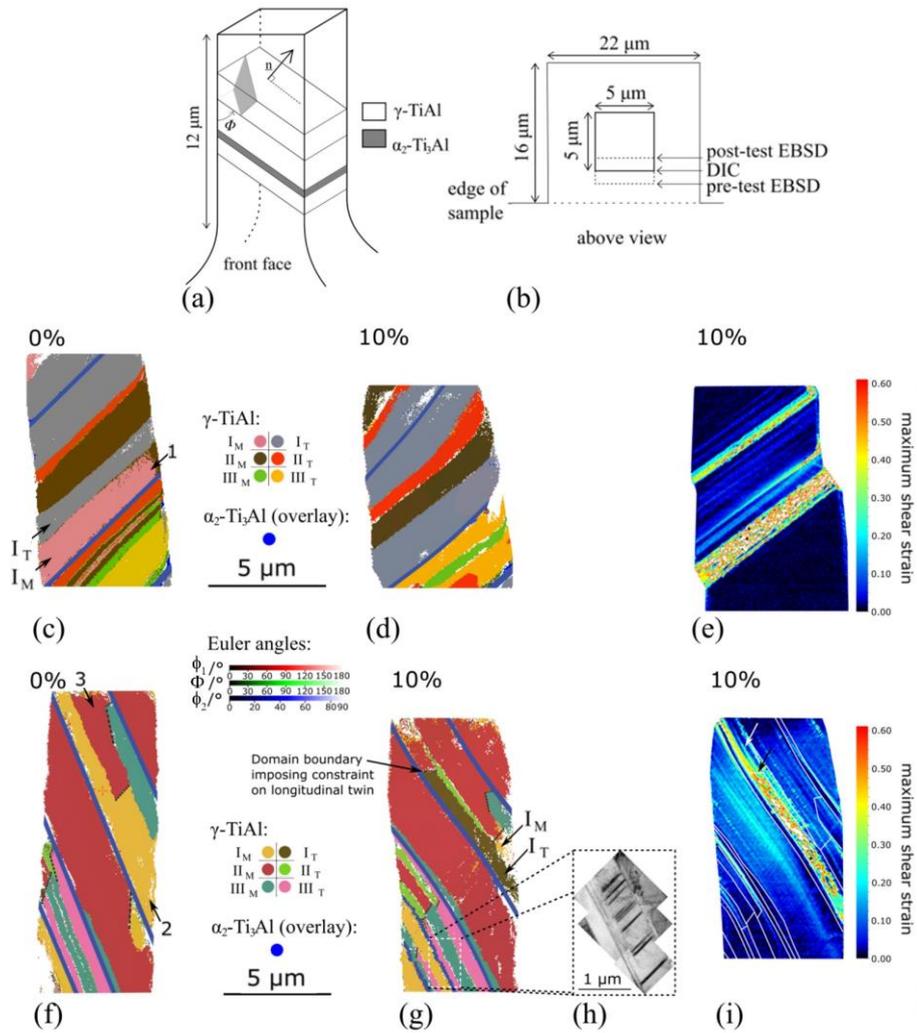

Fig. 2. Experimental analysis of deformation in lamellar TiAl micropillars [16], with geometries in (a, b). Between the before and after compression EBSD maps in (c, d) and (f, g), for the 25° and 55° micropillar, respectively, the occurrence of longitudinal twinning can be identified, as per the arrows numbered 1 to 3. TEM investigation (h) showing transverse mechanical twinning occurred in the $III_T$ $\gamma$ variant in the 25° micropillar. Strain mapping by digital image correlation (e, i) on the same regions of the pillars as for EBSD scans.

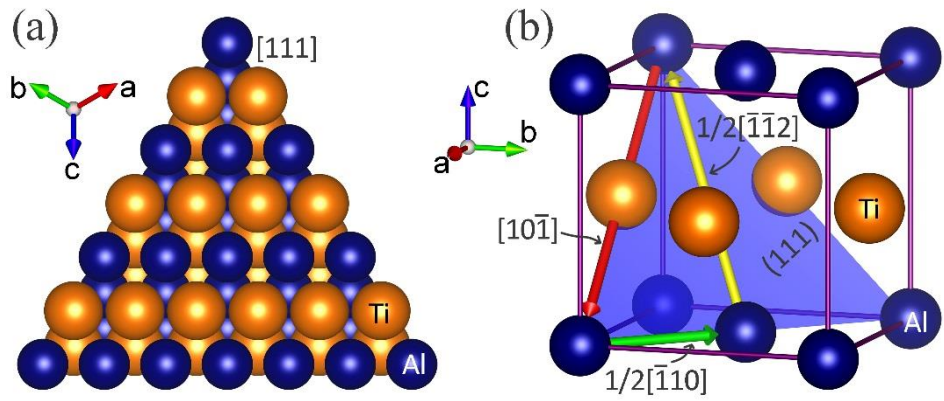

Fig. 3. Atomic structure of γ phase. (a) Close-packed (111) planes stacked by 3 layers. (b) Crystal unit cell and potential deformation systems on (111) plane, including the slip of ordinary $1/2[\bar{1}10]$ and $[10\bar{1}]$ super-dislocations, and unidirectional twinning along $[11\bar{2}]$.

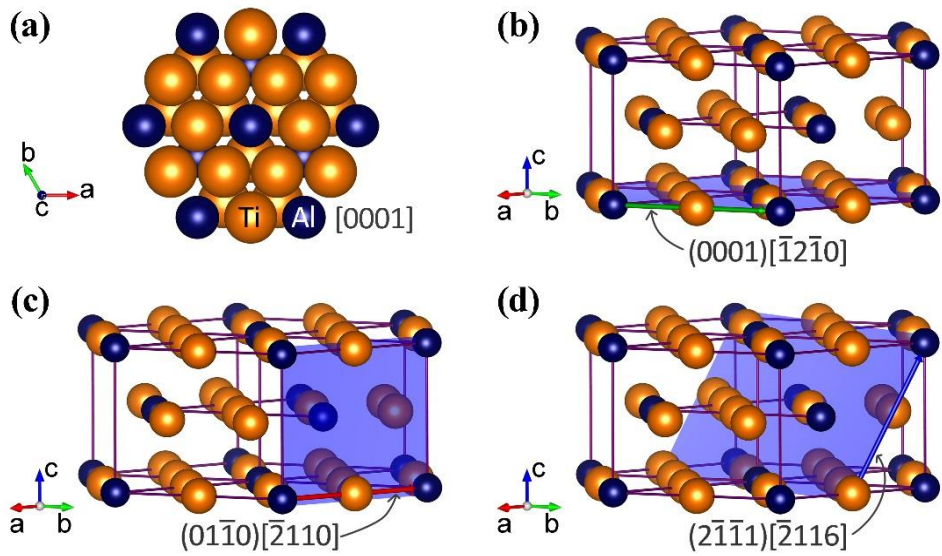

Fig. 4. Atomic structure of α$_2$ phase. (a) Close-packed (0001) planes stacked by 2 layers. Potential slip systems of (b) basal $(0001)[\bar{1}2\bar{1}0]$, (b) prismatic $(01\bar{1}0)[\bar{2}110]$, and (c) pyramidal $(2\bar{1}\bar{1}1)[\bar{2}116]$ slip.

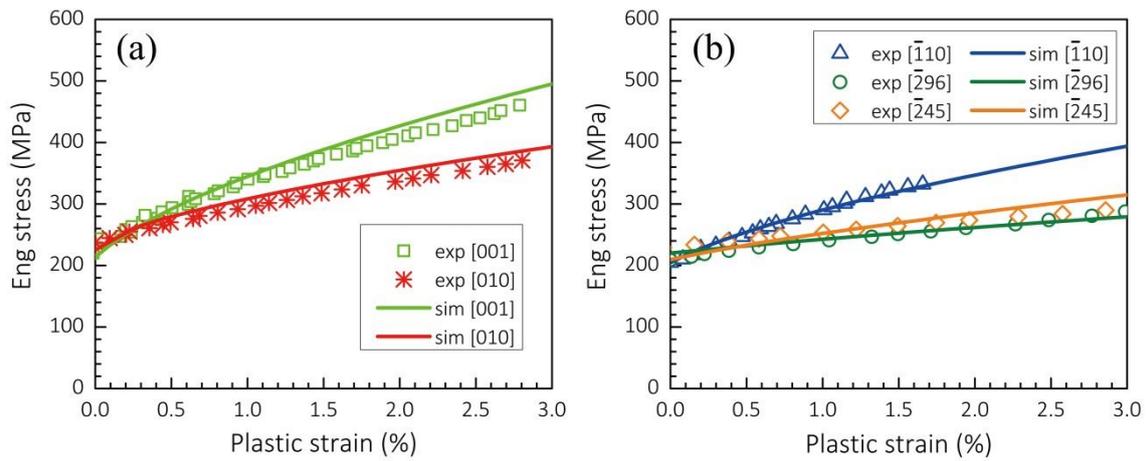

Fig. 5. The engineering stress − plastic strain curves of γ phase obtained from experiments [55] and simulations. (a) Compression along low-indexed orientations which are used to identify the mechanical parameters in model. (b) Compression along high-indexed orientations used the same parameters as those of low-indexed ones.

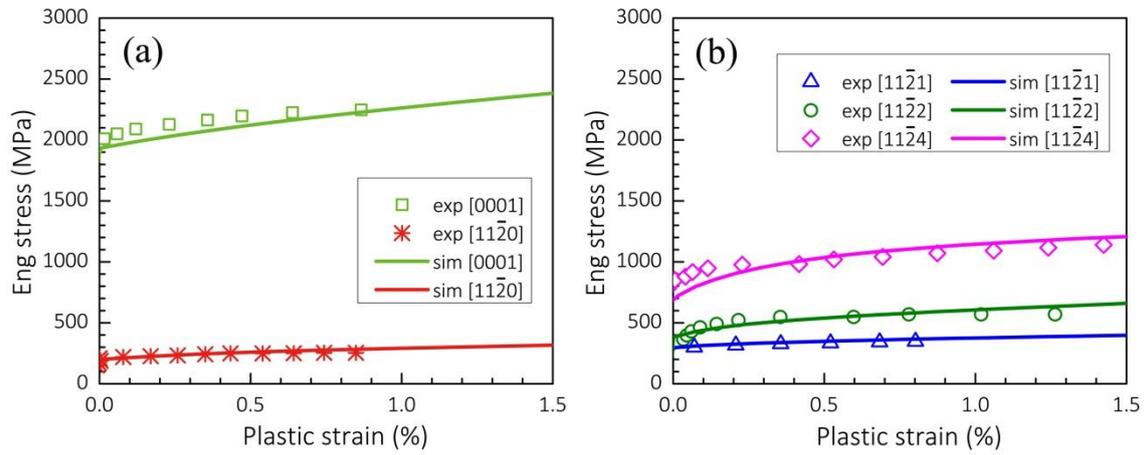

Fig. 6. The engineering stress – plastic strain curves of $\alpha_2$ phase obtained by experiments [56] and simulations. (a) Compressions along low-indexed orientations to identify the parameters in model. (b) Simulated curves with the same parameters as those of low-indexed ones showing good agreement with experimental results.

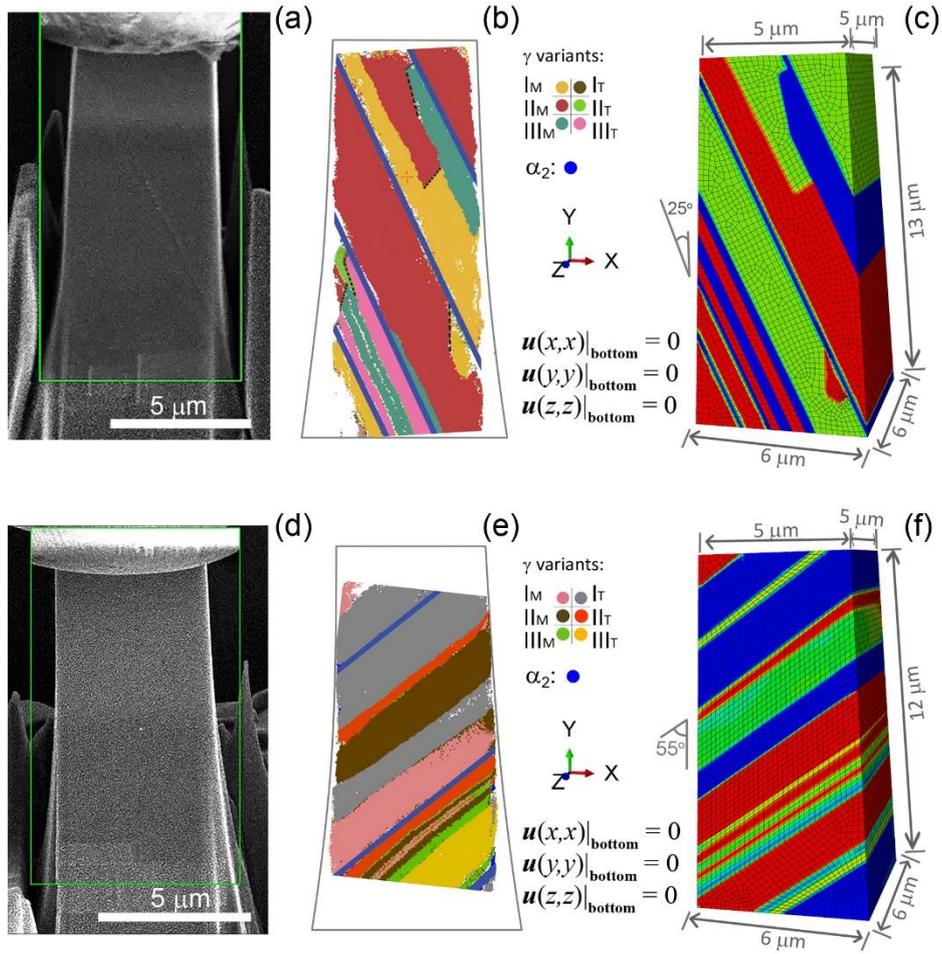

Fig. 7. Geometries of micropillar and corresponding FEM model [16]. (a) The pillar with $\phi = 25^o$ at the beginning of compression. (b) EBSD phase map at the front face showing variants of $\gamma$ phase and $\alpha_2$ lamellae. (c) FEM model based on 7a and 7b. (d-f) The $\phi = 55^o$ micropillar and corresponding EBSD map and FEM model construction.

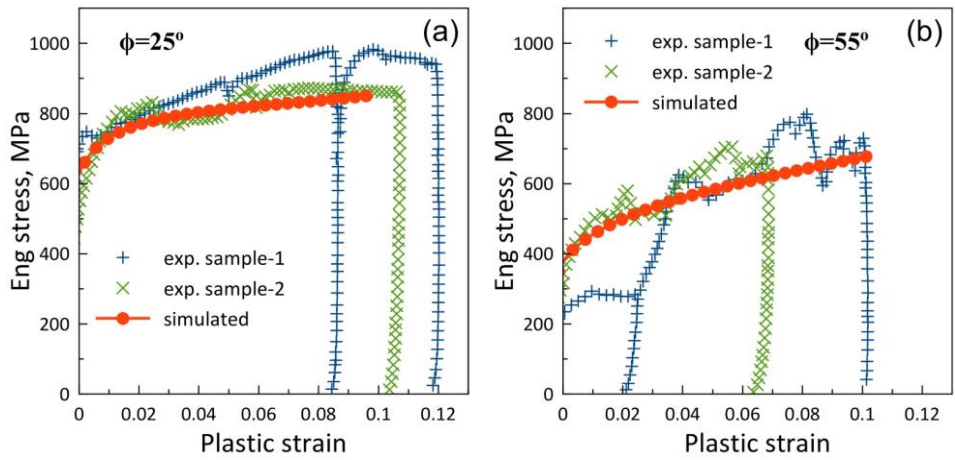

Fig. 8. Experimental and simulated engineering stress – plastic strain curves of (a) $25^o$ micropillar and (b) $55^o$ micropillar showing orientation dependent mechanical properties. Note that EBSD maps (shown in Fig. 7) correspond to the sample 1 of the $25^o$ and $55^o$ pillars.

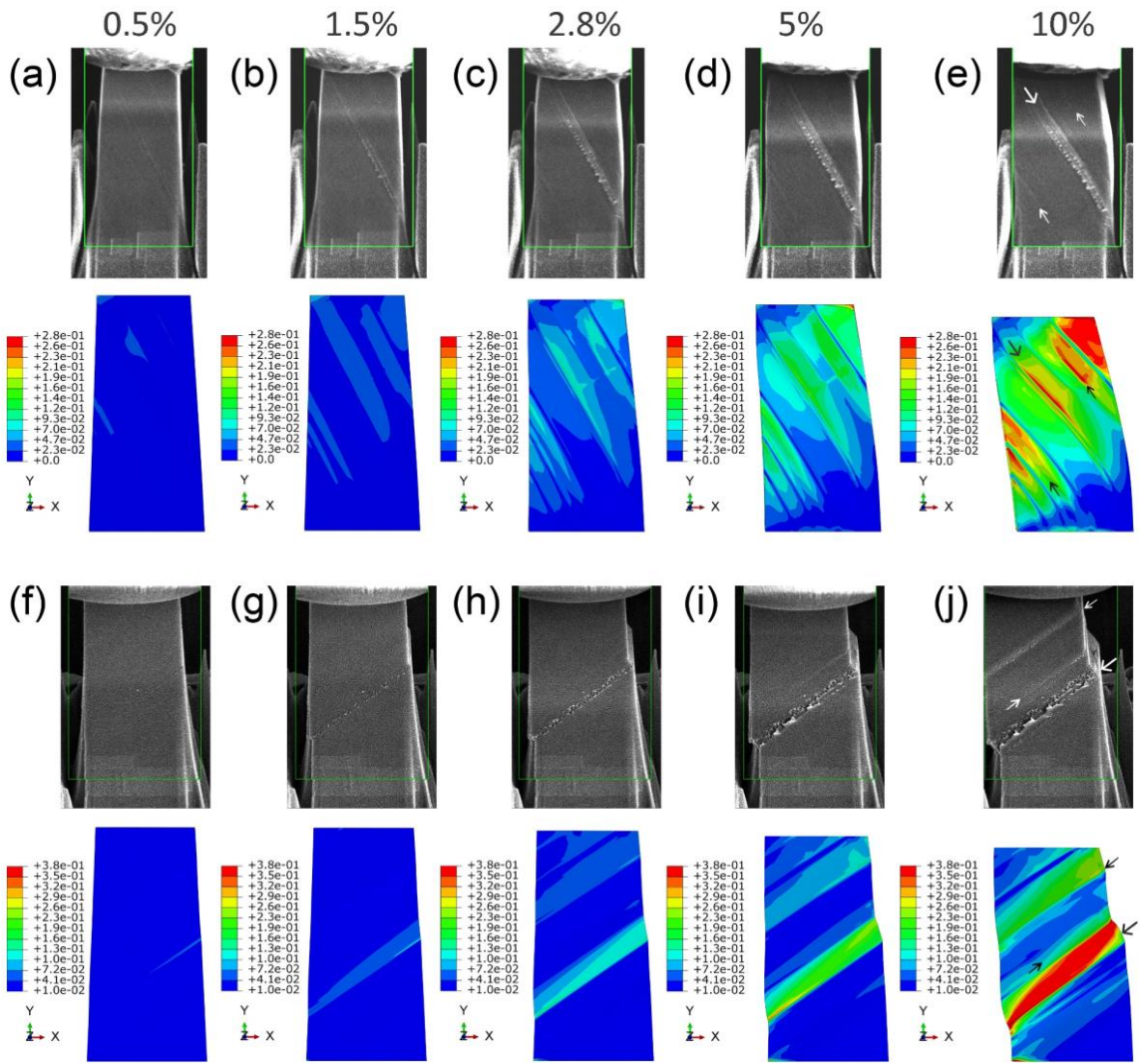

Fig. 9. Compressive deformation of (a – e) 25$^o$ micropillar and (f – j) 55$^o$ micropillar. The plastic strain calculated based on the displacement of the top surface of the micropillars is given at the top of columns 1 – 5. The strain maps show the von Mises strain, and the arrows mark the location of deformation bands (The limits of colour-coded legends were based on the minimum and maximum strains in the two micropillars).

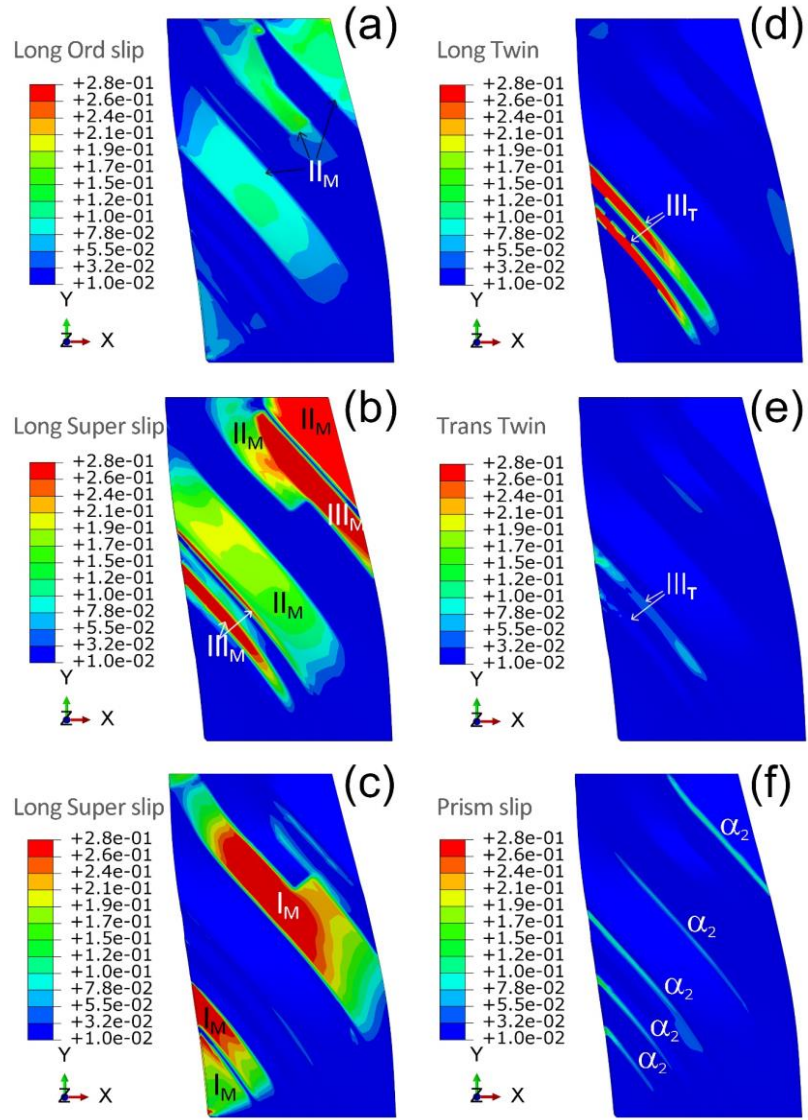

Fig. 10. Accumulated shear strain in individual lamellae associated with the different deformation mechanisms in the $25^o$ micropillar after 10% compression. (a) Longitudinal slip for ordinary dislocations (Long Ord slip) in type II $\gamma$ matrix ($II_M$). (b) Longitudinal slip for superdislocations (Long Super slip) in $II_M$ and type III $\gamma$ matrix ($III_M$). (c) Longitudinal slip for superdislocations in type I $\gamma$ matrix ($I_M$). (d) Longitudinal and (e) Transversal twinning in $\gamma$ type III twin ($III_T$) (Long and Trans Twin). (f) Prismatic slip in $\alpha_2$ lamellae (Prism slip).

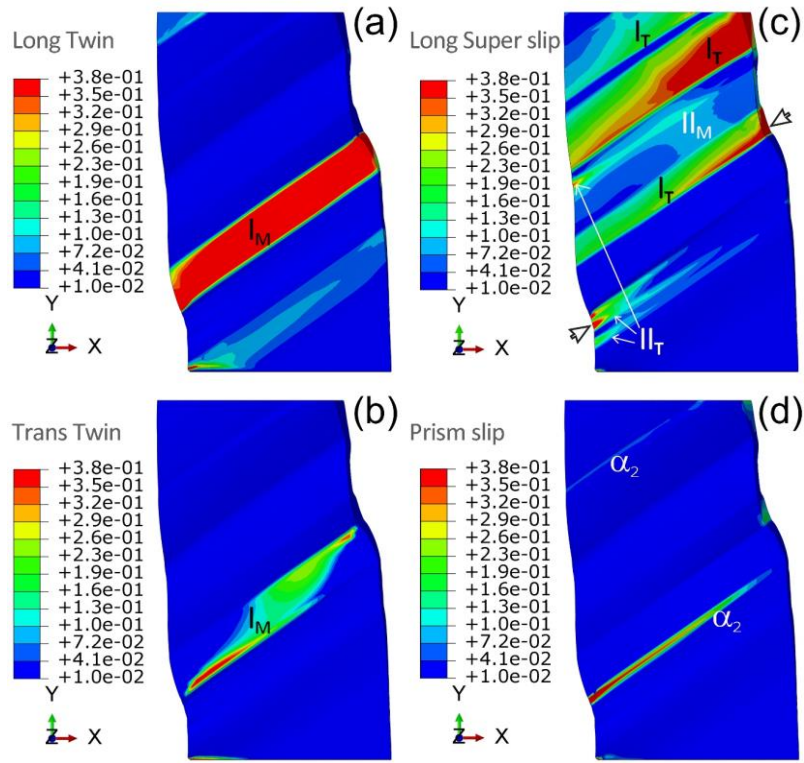

Fig. 11. Accumulated shear strain in individual lamellae from different mechanisms in the $55^o$ micropillar after 10% compression. (a) Longitudinal and (b) transversal twinning in $\gamma$ type I marix ($I_M$) (Long and Trans Twin). (c) Longitudinal slip for superdislocations (Long Super slip) in $\gamma$ type I twin ($I_T$), as well as in some areas of type II matrix ($II_M$) and twin ($II_T$). (d) Prismatic slip in $\alpha_2$ lamellae (Prism slip).

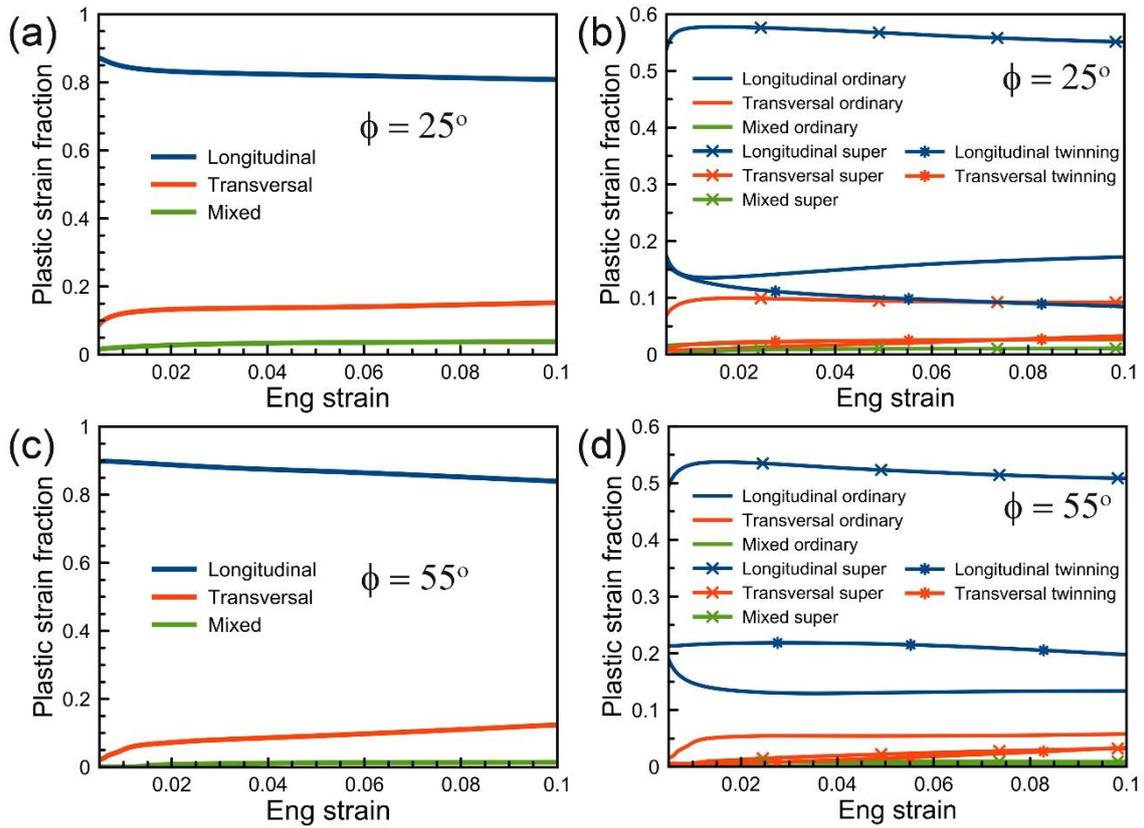

Fig. 12. Plastic strain fraction for different deformation modes and mechanisms in $\gamma$ phase. (a) and (c) The plastic strain fractions categorised on the basis of deformation modes for two pillars, together with (b) and (d) plastic strain fractions differentiated according to the deformation mechanism.

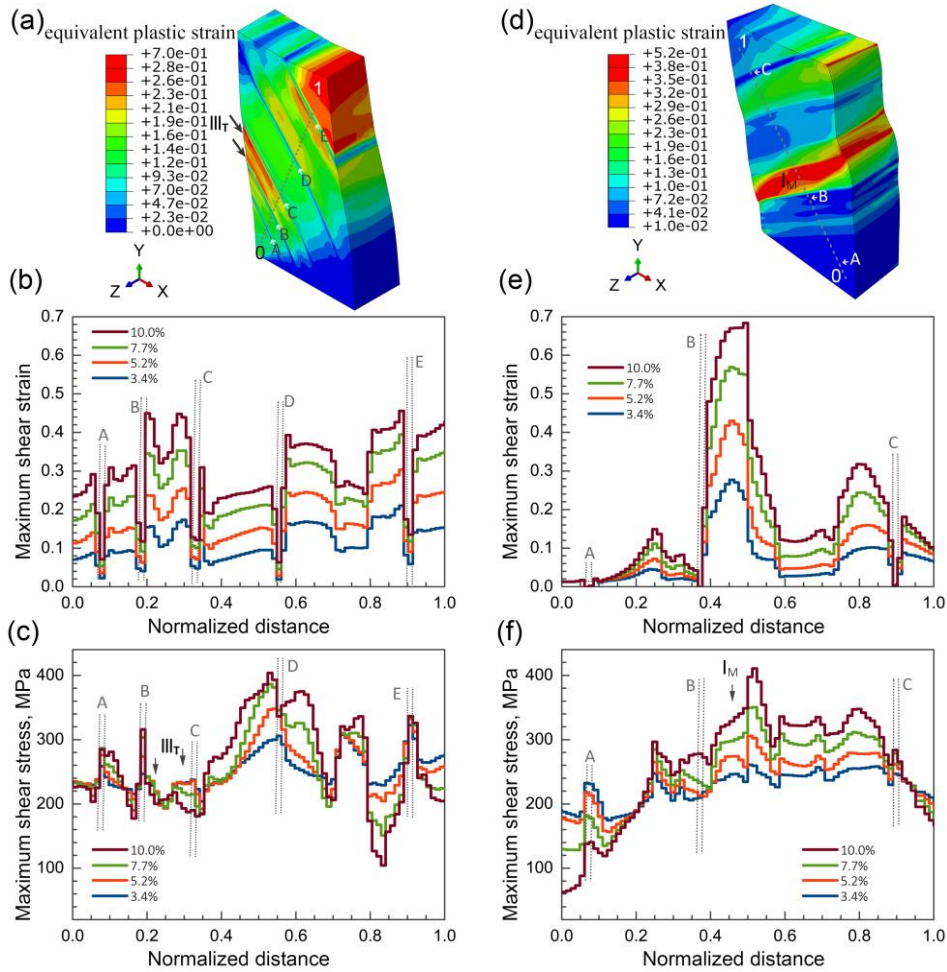

Fig. 13. Inhomogeneous accommodation of plastic deformation between lamellae. (a) Simulated equivalent plastic strain at the centre section of the $25^o$ micropillar after 10% compression. Distribution and evolution of maximum (b) shear strains and (c) shear stresses (on planes parallel with lamellar interface) along the dashed path at different levels of compressive strain. (d)− (f) Similar simulation in the $55^o$ micropillar. Letters A to E indicate $\alpha_2$ lamellae marked by dotted lines in (b,c,e,f).

Table 1.

Slip and twinning rule properties for $\gamma$ and $\alpha_2$ phases at room temperature ($T = 293$ K).

| Parameters | | $\gamma$ phase | | $\alpha_2$ phase |
|---|---|---|---|---|
| $\nu_D$, Hz | | $1.0 \times 10^{11}$ | | |
| $k$, JK$^{-1}$ | | $1.38 \times 10^{-23}$ | | |
| $\Delta F$, J | | $7.8 \times 10^{-20}$ | | |
| $\gamma_0$ | | $6.0 \times 10^{-4}$ | | |
| $\eta$, μm$^{-2}$ | | 460 | | 100 |
| $\rho_m$, μm$^{-2}$ | | 5.0 | | 4.0 |
| $\rho_{im}$, μm$^{-2}$ | | 1.0 | | 1.0 |
| | *ordinary* | $2.83 \times 10^{-4}$ | *basal* | $5.77 \times 10^{-4}$ |
| $b$, μm | *super* | $5.71 \times 10^{-4}$ | *prism* | $5.77 \times 10^{-4}$ |
| | *twinning* | $1.65 \times 10^{-4}$ | *pyram* | $5.44 \times 10^{-4}$ |

Table A.1.

Slip and twinning systems and corresponding deformation modes in $\gamma$ phase, where $k_{so} = 1.1$ and $\tau_0 = 45$ MPa for single phase samples, while $k_{so} = 1.8$ and $\tau_0 = 70$ MPa for lamellar crystals. The $(1\bar{1}1)$ plane is parallel with the lamellar boundaries.

| No. | Slip systems | Initial CRSS | Mechanism | Deformation mode |
|---|---|---|---|---|
| 1 | $(111)[1\bar{1}0]$ | $\tau_0$ | Ordinary | Transversal |
| 2 | $(111)[01\bar{1}]$ | $k_{so} \cdot \tau_0$ | Super | Transversal |
| 3 | $(111)[10\bar{1}]$ | $k_{so} \cdot \tau_0$ | Super | Mixed |
| 4 | $(\bar{1}11)[110]$ | $\tau_0$ | Ordinary | Mixed |
| 5 | $(\bar{1}11)[01\bar{1}]$ | $k_{so} \cdot \tau_0$ | Super | Transversal |
| 6 | $(\bar{1}11)[101]$ | $k_{so} \cdot \tau_0$ | Super | Transversal |
| 7 | $(1\bar{1}1)[110]$ | $\tau_0$ | Ordinary | Longitudinal |
| 8 | $(1\bar{1}1)[011]$ | $k_{so} \cdot \tau_0$ | Super | Longitudinal |
| 9 | $(1\bar{1}1)[10\bar{1}]$ | $k_{so} \cdot \tau_0$ | Super | Longitudinal |
| 10 | $(11\bar{1})[1\bar{1}0]$ | $\tau_0$ | Ordinary | Mixed |
| 11 | $(11\bar{1})[011]$ | $k_{so} \cdot \tau_0$ | Super | Transversal |
| 12 | $(11\bar{1})[101]$ | $k_{so} \cdot \tau_0$ | Super | Transversal |
| 13 | $(111)[11\bar{2}]$ | $\tau_0$ | Twinning | Transversal |
| 14 | $(\bar{1}11)[\bar{1}1\bar{2}]$ | $\tau_0$ | Twinning | Transversal |
| 15 | $(1\bar{1}1)[1\bar{1}\bar{2}]$ | $\tau_0$ | Twinning | Longitudinal |
| 16 | $(11\bar{1})[112]$ | $\tau_0$ | Twinning | Transversal |

Table A.2.

Slip systems in hexagonal $\alpha_2$ lamellae, where $\tau_p = 85$ MPa, $k_{bp} = 2.0$, $k_{pyp} = 9.5$, as calibrated from the simulation of single phase crystals, which were also employed in simulation of lamellar crystals. The (0001) plane is parallel with lamellar interfaces.

| No. | Slip systems | Initial CRSS | Mechanism | Deformation mode |
|---|---|---|---|---|
| 1 | $(0001)[2\bar{1}\bar{1}0]$ | $k_{bp} \cdot \tau_p$ | Basal | Longitudinal |
| 2 | $(0001)[\bar{1}2\bar{1}0]$ | $k_{bp} \cdot \tau_p$ | Basal | Longitudinal |
| 3 | $(0001)[\bar{1}\bar{1}20]$ | $k_{bp} \cdot \tau_p$ | Basal | Longitudinal |
| 4 | $(10\bar{1}0)[\bar{1}2\bar{1}0]$ | $\tau_p$ | Prismatic | Mixed |
| 5 | $(01\bar{1}0)[\bar{2}110]$ | $\tau_p$ | Prismatic | Mixed |
| 6 | $(\bar{1}100)[\bar{1}\bar{1}20]$ | $\tau_p$ | Prismatic | Mixed |
| 7 | $(11\bar{2}1)[\bar{1}\bar{1}26]$ | $k_{pyp} \cdot \tau_p$ | Pyramidal | Transversal |
| 8 | $(1\bar{2}11)[\bar{1}2\bar{1}6]$ | $k_{pyp} \cdot \tau_p$ | Pyramidal | Transversal |
| 9 | $(\bar{2}111)[2\bar{1}\bar{1}6]$ | $k_{pyp} \cdot \tau_p$ | Pyramidal | Transversal |
| 10 | $(\bar{1}\bar{1}21)[11\bar{2}6]$ | $k_{pyp} \cdot \tau_p$ | Pyramidal | Transversal |
| 11 | $(\bar{1}2\bar{1}1)[1\bar{2}16]$ | $k_{pyp} \cdot \tau_p$ | Pyramidal | Transversal |
| 12 | $(2\bar{1}\bar{1}1)[\bar{2}116]$ | $k_{pyp} \cdot \tau_p$ | Pyramidal | Transversal |

Table A.3.

The maximum Schmid factors for variants in the γ phase for different deformation mechanisms; Suffixes C and T refer to the twinning behaviour activated under compression and tension, respectively, while the abbreviated letters Ord, Sup and Twn denote ordinary slip, superdislocation slip and twinning, respectively.

| $\phi$ | Variant | Longitudinal | | | Mixed | | | Transversal | | |
|---|---|---|---|---|---|---|---|---|---|---|
| | | Ord | Sup | Twn | Ord | Sup | Twn | Ord | Sup | Twn |
| 25º | $I_M$ | 0.28 | 0.35 | 0.25C | 0.29 | 0.12 | -- | 0.00 | 0.37 | 0.26C |
| | $I_T$ | 0.28 | 0.36 | 0.25T | 0.48 | 0.35 | -- | 0.28 | 0.35 | 0.25T |
| | $II_M$ | 0.36 | 0.28 | 0.12C | 0.11 | 0.17 | -- | 0.17 | 0.38 | 0.39T |
| | $II_T$ | 0.36 | 0.28 | 0.12T | 0.36 | 0.13 | -- | 0.36 | 0.48 | 0.35T |
| | $III_M$ | 0.08 | 0.36 | 0.37T | 0.18 | 0.29 | -- | 0.37 | 0.16 | 0.13C |
| | $III_T$ | 0.08 | 0.36 | 0.36C | 0.13 | 0.48 | -- | 0.12 | 0.48 | 0.48C |
| 55º | $I_M$ | 0.10 | 0.45 | 0.46C | 0.23 | 0.23 | -- | 0.15 | 0.49 | 0.43C |
| | $I_T$ | 0.10 | 0.44 | 0.45T | 0.10 | 0.06 | -- | 0.12 | 0.46 | 0.47T |
| | $II_M$ | 0.44 | 0.34 | 0.13T | 0.24 | 0.26 | -- | 0.49 | 0.39 | 0.32T |
| | $II_T$ | 0.44 | 0.35 | 0.15C | 0.46 | 0.04 | -- | 0.05 | 0.36 | 0.15C |
| | $III_M$ | 0.34 | 0.44 | 0.31T | 0.39 | 0.03 | -- | 0.22 | 0.49 | 0.41T |
| | $III_T$ | 0.34 | 0.44 | 0.32C | 0.05 | 0.10 | -- | 0.36 | 0.46 | 0.32C |